\documentclass{llncs}

\usepackage{url}

\usepackage{tabularx}
\usepackage{multirow}
\usepackage{color}
\usepackage{booktabs}
\usepackage[square,comma,numbers,sort&compress,sectionbib]{natbib}
\usepackage{enumerate}
\usepackage{enumitem}

\usepackage{tcolorbox}
\usepackage{xspace}
\usepackage{amssymb}
\usepackage{xcolor}
\usepackage{soul}
\usepackage{subcaption}
\usepackage{graphicx}

\definecolor{varblue}{RGB}{0, 102, 204}
\definecolor{vargreen}{RGB}{0, 153, 0}

\newcommand{\dataset}{SciNUP\xspace}

\begin{document}
\mainmatter

\title{SciNUP: Natural Language User Interest Profiles for Scientific Literature Recommendation}

\author{Mariam Arustashvili
\and Krisztian Balog
}
\authorrunning{Arustashvili and Balog} 
\tocauthor{Mariam Arustashvili, Krisztian Balog}
\institute{University of Stavanger, Stavanger, Norway,\\
\email{\{mariam.arustashvili, krisztian.balog\}@uis.no}}

\maketitle

\begin{abstract}
The use of natural language (NL) user profiles in recommender systems offers greater transparency and user control compared to traditional representations. 
However, there is scarcity of large-scale, publicly available test collections for evaluating NL profile-based recommendation. 
To address this gap, we introduce SciNUP, a novel synthetic dataset for scholarly recommendation that leverages authors' publication histories to generate NL profiles and corresponding ground truth items. We use this dataset to conduct a comparison of baseline methods, ranging from sparse and dense retrieval approaches to state-of-the-art LLM-based rerankers.
Our results show that while baseline methods achieve comparable performance, they often retrieve different items, indicating complementary behaviors. At the same time, considerable headroom for improvement remains, highlighting the need for effective NL-based recommendation approaches.
The SciNUP dataset thus serves as a valuable resource for fostering future research and development in this area.

\end{abstract}

\section{Introduction}

Natural language (NL) user profiles have recently emerged as a promising para\-digm in recommender systems~\citep{Balog:2019:SIGIR,Penaloza:2025:WWW,Ramos:2024:ACL}, offering an interpretable and flexible alternative to traditional user representations, such as embeddings derived from interaction histories~\citep{Radlinski:2022:SIGIR}. By aligning with the way humans naturally describe their interests and preferences, NL profiles enable greater transparency and user agency: users can inspect, understand, and edit their profiles in plain language. NL profile-based systems move beyond pure behaviorism, effectively scaling participatory design and  giving users more control over the recommendation process~\cite{Ekstrand:2016:RecSys}, thereby shifting the focus from recommending for the users to recommending with them~\cite{Ekstrand:2025:RecSys}. This makes NL user profiles highly suitable for applications prioritizing explainability, personalization, and user control.

Despite the growing interest in NL-based user representations, a significant gap remains in the current research landscape: the scarcity of large-scale, publicly available test collections specifically designed to evaluate models and algorithms for NL profile-based recommendation~\cite{Ramos:2024:ACL}. 
Creating such datasets is challenging because soliciting natural language interest descriptions from users and collecting relevance judgments for recommendations is a time-intensive and expensive process that is difficult to scale~\citep{Sanner:2023:RecSys}.
Consequently, many existing approaches rely on inferring NL user profiles from readily available user interaction data like reviews~\cite{Ramos:2024:ACL, Mysore:2023:RecSys} or ratings~\cite{Penaloza:2025:WWW,Gao:2025:arXiv, Zhou:2024:arXiv}. Furthermore, existing studies are often limited in domain, predominantly focusing on movie or book recommendations~\citep{Penaloza:2025:WWW,Ramos:2024:ACL,Sanner:2023:RecSys,Gao:2025:arXiv}.  
While some work, such as \citep{Mysore:2023:SIGIR}, explores scientific paper recommendation, the use of NL user profiles for this domain has not been explored to date, nor are there public datasets to support it. With this work, we aim to bridge this gap.

\begin{figure}[t]
    \ttfamily
    \scriptsize
    \begin{tabular}{p{0.98\linewidth}}
    \toprule
    My research lies at the intersection of information theory, probability, and convex geometry, with a particular focus on entropy inequalities, information-theoretic limits, and log-concave distributions. I investigate fundamental properties of entropy in both discrete and continuous settings, develop analogs of additive combinatorics in an information-theoretic framework, and explore applications to problems such as compound Poisson approximation, polar coding, and combinatorial geometry. My work often uncovers deep connections between probabilistic methods, functional inequalities, and structural aspects of high-dimensional spaces. \\
    \bottomrule
    \end{tabular}
    \vspace*{-0.75\baselineskip}
    \caption{Example NL user interest profile from our dataset.}
    \label{fig:example}
    \vspace{-1\baselineskip}
\end{figure}

We introduce \dataset (\textbf{Sci}entific \textbf{N}atural Language \textbf{U}ser \textbf{P}rofiles), a novel large-scale synthetic dataset for NL-based recommendation. Our approach simulates a realistic scholarly recommendation scenario: for a given individual, we leverage their authored papers, split temporally, to create an NL user interest profile from their earlier works by prompting a large language model (LLM). This NL profile, illustrated in Fig.~\ref{fig:example}, then serves as the basis for predicting future research interests, with ground truth derived from references within their later publications. Each NL profile is associated with a set of candidate items, enabling a fair comparison across different recommender approaches. We employ multiple LLMs and prompts for profile generation to reduce model-specific biases and enhance the generality of the resulting profiles. Additionally, NL profiles are automatically classified based on the breadth of user interest into narrow, medium, and broad categories, enabling further analysis and performance breakdown.
Together, these components constitute a rich test collection specifically designed to enable repeatable and reproducible offline experimentation with NL-based recommender systems in the scholarly domain.

We evaluate a range of methods, from sparse and dense retrieval to LLM-based reranking. Our experiments show a nuanced landscape: while advanced dense and LLM-based methods ultimately set the state of the art, traditional sparse retrieval proves to be a highly competitive baseline that is difficult to surpass. We find that the strengths of these approaches are complementary, with each model excelling on different subsets of users. To highlight this, our simple ensemble of these models achieves the best overall performance by a large margin. This indicates that considerable headroom for improvement remains, establishing our dataset as a valuable benchmark to foster future research and model development.

In summary, the main contributions of this work are twofold: 
\begin{enumerate}[label=(\arabic*)]
    \item We introduce \dataset, a large-scale dataset for NL-based recommendation in the scholarly domain.
    \item We provide a comprehensive performance analysis of diverse recommender approaches on \dataset, from traditional sparse and dense retrieval to state-of-the-art LLM-based reranking, establishing strong baselines for future work.
\end{enumerate}
All resources, including the dataset, baseline methods, and evaluation results, are made publicly available at \url{https://github.com/iai-group/SciNUP}.

\section{Related work}
\label{sec:related}

Editable user profiles, allowing users to inspect, validate, and modify system beliefs about them, have gained traction in recommender systems. 
Early work by \citet{Balog:2019:SIGIR} introduced a template-based method for summarizing user profiles in natural language, relying on keyword or tags.
Later, \citet{Radlinski:2022:SIGIR} proposed the idea of NL user profiles as an alternative to traditional user embedding representations for an intermediate step between user behavior data and item recommendations. 
Building on this idea, \citet{Sanner:2023:RecSys} demonstrated that NL-based recommendations can be competitive with item-based collaborative filtering, particularly in cold-start scenarios.
More recently, studies have increasingly leveraged LLM prompting for NL profile generation based on user reviews, ratings, and/or history~\citep{Gao:2025:arXiv,Penaloza:2025:WWW,Ramos:2024:ACL,Zhou:2024:arXiv, Gagliano:2025:MuRS, Sguerra:2025:RecSys}. For instance, \citet{Penaloza:2025:WWW} provide specific instructions in the prompt as to what aspects to summarize, while \citet{Ramos:2024:ACL} perform preference extraction and ranking before prompting. \citet{Gao:2025:arXiv} optimize the profile encoder through reinforcement learning.
NL profiles are usually evaluated only intrinsically, using recommendation performance as a
proxy for quality~\cite{Gao:2025:arXiv, Penaloza:2025:WWW,Zhou:2024:arXiv}. 
\citet{Ramos:2024:ACL} conducted a user study to assess NL profiles in terms of fluency, informativeness, conciseness, and relevance.
A recent user study in the music recommendation domain shows a weak correlation between the perceived representativeness of the NL profile and recommendation performance~\cite{Sguerra:2025:RecSys}.
Recommendations from NL profiles are typically generated using retrieval-based methods that rank items by similarity in a shared embedding space~\citep{Gao:2025:arXiv, Ramos:2024:ACL}, while others use neural models trained for rating or choice prediction~\citep{Penaloza:2025:WWW, Zhou:2024:arXiv}. 

The task of narrative-driven recommendation~\citep{Bogers:2017:RecSys,Mysore:2023:RecSys} is closely related to NL profile-based recommendation, yet a key distinction lies in their scope. Narratives are inherently more situational and context-dependent (e.g., suggesting music for a specific mood on an autumn afternoon) and typically incorporate explicit positive or negative item examples. NL profiles, conversely, aim to capture a more stable and general set of a person's preferences within a specific domain (e.g., an individual's overall music taste).

Most existing work focuses on the domains of movie~\cite{Gao:2025:arXiv, Zhou:2024:arXiv,Penaloza:2025:WWW, Ramos:2024:ACL}, book~\cite{Gao:2025:arXiv,Penaloza:2025:WWW}, 
or point-of-interest~\cite{Ramos:2024:ACL,Mysore:2023:RecSys,Afzali:2021:SIGIR} recommendation. 
While \citet{Mysore:2023:SIGIR} uniquely addresses scientific paper recommendation, their approach relies on a set of NL concepts, more akin to tag-based methods, and evaluates on sparse (CiteULike) or private (OpenReview) datasets.
In contrast, we generate full-text NL summaries, enabling more expressive and interpretable user representations aligned with the common understanding of NL profiles. 
In the realm of scientific literature recommendation, arXiv is a widely used preprint service, offering a rich dataset~\citep{Clement:2019:arXiv}. 
Services like ArXivDigest \citep{Gingstad:2020:CIKM}, while providing valuable living lab environments, typically assume explicit keyword-based interest expressions or implicit interactions. Our use of NL user profiles represents a significant step towards addressing the gap in explicit user modeling within this domain \citep{Kreutz:2022:IJDL}.

\section{Problem Statement}
\label{sec:problem}

\emph{Natural language user interest profile generation} (\emph{NL profile generation} for short) is the task of generating a natural language description $d_u$ based on a set of items the user has interacted with.
The generated profile should be a concise, fluent, and effective characterization of the user's interests, allowing for interpretability and human understanding while being useful for recommendation~\citep{Radlinski:2022:SIGIR}.

Given a natural language description $d_u$ characterizing the interest of user $u$, \emph{natural language profile-based recommendation} is the task of recommending items from a pool of candidate items $I$ that match the interests of that person. This recommendation task is approached as a ranking problem, where the goal is to produce an ordered list of candidate items such that items more relevant to the user's interests appear higher in the list.

We address these two tasks---NL profile generation and NL profile-based recommendation---in turn in the following two sections, specifically in the context of scientific literature recommendation.

\section{The \dataset Dataset}
\label{sec:dataset}

We present \dataset (\textbf{Sci}entific \textbf{N}atural Language \textbf{U}ser \textbf{P}rofiles), a synthetic dataset  designed for evaluating natural language-based recommendation in the scholarly domain. 
We simulate a real-world scenario in which a researcher---hereafter referred to as the \emph{user}---is looking for article recommendations relevant to their specific expertise and interests. To model this authentically, we leverage the user's publication history. We utilize their past authored papers to create NL user interest profiles; subsequently, the cited references in their later publications serve as the ground truth for evaluating recommendation effectiveness.

\subsection{Approach}
\label{sec:dataset:approach}

\begin{figure}[t]
    \centering
    \includegraphics[width=1\linewidth]{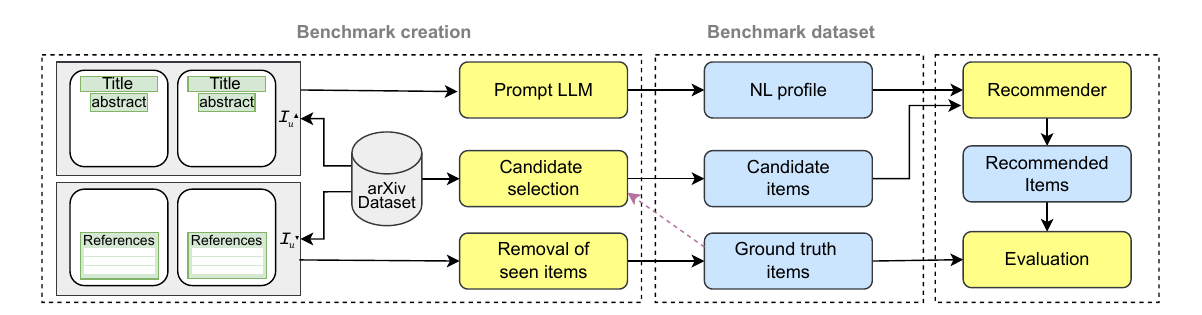}
    \caption{Overview of our approach.}
    \label{fig:pipeline}
\end{figure}

The construction of our benchmark dataset, illustrated in Fig.~\ref{fig:pipeline}, centers on three key components: NL profiles, candidate items, and ground truth. Our methodology begins by sampling users (authors) from a scholarly dataset (arXiv).  
For each user, their published papers are temporally divided into an earlier and a later portion. The \emph{earlier portion} is utilized for generating an NL interest profile through LLM prompting, while the \emph{later portion} is utilized for creating the ground truth set of recommendations. For each user, we also sample additional papers from arXiv to create a set of candidate items. The subsequent recommendation task involves ranking these candidate papers with respect to the NL profile. Further details on each step are provided below.

\subsubsection{\textbf{Scholarly Dataset}}
We use the public arXiv dataset released by \citet{Clement:2019:arXiv}.\footnote{\url{https://github.com/mattbierbaum/arxiv-public-datasets/releases/tag/v0.2.0}} To ensure maximum accessibility and reproducibility, we rely exclusively on article metadata, which is released under a CC0 license. 
This metadata includes the arXiv paper ID, author list, title, abstract, version, and category listings for 1,506,500 articles.
Furthermore, we integrate a citation network provided by the same source to establish relational links between papers.
The dataset covers articles published up to March 1, 2019. We performed author name normalization, assigned unique user IDs, and retained only users with a valid name who had published at least 10 papers.
We only kept articles that had metadata and were also present in the citation network.

\subsubsection{\textbf{NL Profile Generation}}

We sample $n{=}1,000$ users from the filtered arXiv dataset. 
For each user $u$, let $\mathcal{I}_u$ denote the articles they authored. We temporarily order $\mathcal{I}_u$ and divide to earlier portion, $\mathcal{I}_u^\blacktriangle$ (for NL profile construction), and later portion, $\mathcal{I}_u^\blacktriangledown$ (for ground truth creation). 
For simplicity, we divide $\mathcal{I}_u$ in half to ensure that there are enough articles for both NL profile generation and ground truth creation; exploring other division strategies is left for future work.

The NL profile descriptions are generated by prompting an LLM and providing the list of articles in $\mathcal{I}_u^\blacktriangle$, represented by their titles and abstract, as context and input to the prompt. 
To ensure generalizability, we designed two different prompt versions (A and B), shown in Fig.~\ref{fig:prompts}. Prompt A provides more details about the purpose of the summary and specifies the desired length, while Prompt B leaves the LLM more freedom. Furthermore, we utilized two different LLMs: Llama-4 Maverick and GPT-4o. The author set is split uniformly into four groups (250 authors each), corresponding to the possible prompt-LLM combinations. 

\begin{figure}[!t]
    \ttfamily
    \scriptsize
    \begin{tabular}{p{0.98\linewidth}}
    \toprule
        Below are the titles and abstracts of scientific papers I have authored. 
        Based on this information, generate a concise \\
        \textcolor{vargreen}{$\langle$IF Prompt A$\rangle$} \\
        first-person research profile that summarizes my main research interests and areas of expertise. \\
        The profile should be written in no more than three sentences and should clearly identify the key themes, research trends, and topics I focus on. The purpose of this profile is to support a scientific literature recommendation system, so it should accurately reflect my research focus to help match me with relevant publications.\\
        \textcolor{vargreen}{$\langle$ELIF Prompt B$\rangle$} \\
        description of my research interests, characterizing the key topics and areas of expertise. \\
        \textcolor{vargreen}{$\langle$/IF$\rangle$} \\
        \\
        This is the list of my publications: \\
        \textcolor{varblue}{[List of articles (titles and abstracts)]} \\
        \\
        Write the profile in first person. Return nothing but the generated profile. \\
    \bottomrule
    \end{tabular}
    \caption{Prompt template used for NL profile generation. Placeholders for variables are in \textcolor{varblue}{[square brackets]}, while conditional logic is marked by \textcolor{vargreen}{$\langle$IF$\rangle$}...\textcolor{vargreen}{$\langle$/IF$\rangle$}.}
    \label{fig:prompts}
\end{figure}

\begin{figure}[!t]
    \ttfamily
    \scriptsize
    \begin{tabular}{p{0.98\linewidth}}
    \toprule
        You are an expert in classifying scholarly user interest profiles. Your task is to analyze a given natural language description of a user's research interests and classify it as 'Narrow', 'Medium', or 'Broad' based on the specificity and scope of the topics mentioned. \\
        \\
        Here are the definitions for each category: \\
        \\
        Narrow: The profile describes highly specific interests within a single, well-defined subfield. The language is often technical and domain-specific. \\
        Medium: The profile covers a single, broader field or several related topics. The interests are connected but not as specific as a narrow profile. \\
        Broad: The profile covers a wide range of disparate topics or a very general field. The interests may not be directly connected. \\
        \\
        Examples: \\
        \\
        User Profile: 'My research focuses on the optimization of federated learning algorithms for on-device natural language processing, specifically for low-resource languages.' \\
        Classification: Narrow \\
        \\
        User Profile: 'I am interested in the intersection of artificial intelligence and medicine. My work involves using computer vision for medical image analysis and developing predictive models for disease progression using electronic health records.' \\
        Classification: Medium \\
        \\
        User Profile: 'I have a passion for technology and its role in society. I'm interested in everything from robotics and human-computer interaction to the ethical implications of AI and the future of work. I also enjoy historical perspectives on technological innovation.' \\
        Classification: Broad \\
        \\
        Please classify the following user profile. Return only one word, Narrow, Medium or Broad. \\
        \\
        User Profile: \textcolor{varblue}{[nl\_profile]} \\
        Classification: \\
    \bottomrule
    \end{tabular}
    \caption{Prompt template used for NL profile classification. Placeholders for variables are in \textcolor{varblue}{[square brackets]}.}
    \label{fig:classification-prompt}
\end{figure}

%
%

\subsubsection{\textbf{Ground Truth}}

The later portion of the articles, $\mathcal{I}_u^\blacktriangledown$, which were not used for NL profile generation, are utilized for ground truth creation.  We consider all papers referenced by any article in $\mathcal{I}_u^\blacktriangledown$ as ground truth items, provided they were not ``seen'' by the user before, that is, referenced by any of the papers in $\mathcal{I}_u^\blacktriangle$ or authored by the individual themselves (i.e., exclude self-citations). 
We denote this set of ground truth items as $\mathcal{G}_u$. 
In this initial approach, we do not distinguish between the varying degrees of relevance among these items.

\subsubsection{\textbf{Candidate Selection}}

For each user $u$, we construct a unique set of candidate items, $\mathcal{C}_u$, which serve as the pool of articles from which recommendations will be made. 
We set the number of candidate items $m=1000$ for each user. 
For simplicity, we employ a modified version of the One-Plus-Random methodology~\citep{Bellogin:2011:RecSys}, which we term \emph{N-plus-random}: $N$ items are drawn from the ground truth set ($\mathcal{G}_u$), while the remaining ($m{-}N$) items are randomly sampled from the arXiv dataset.
These random samples are selected from papers within the same categories as those previously authored by the user. The sampling process is weighted: categories where a user has published more receive a higher probability of contributing candidate items. Before sampling, all ``seen'' items (previously cited or self-authored, as above) are excluded from the pool of available articles.

\begin{table}[t]
\centering
\caption{Examples of Narrow, Medium, and Broad natural language research profiles from the \dataset dataset.}
\label{tab:breadth_examples}

\renewcommand{\arraystretch}{1.25}
\setlength{\tabcolsep}{6pt}

\begin{tabularx}{\textwidth}{p{0.02\textwidth} X}
\toprule
\multirow{1}{*}[-1.2em]{\rotatebox{90}{\textbf{Narrow}}} &
\scriptsize \ttfamily
My research focuses on precision measurements in flavor physics, particularly the semileptonic and rare decays of B and D mesons, to test the Standard Model and probe for signs of new physics. I specialize in the determination of CKM matrix elements, lepton flavor universality tests, and angular analyses of rare decays, using large datasets collected by the Belle detector at the KEKB collider. I am also involved in searches for rare charm decays and studies of bottomonium spectroscopy to investigate hadronic decay dynamics. \\
\midrule
\multirow{1}{*}[-1.2em]{\rotatebox{90}{\textbf{Medium}}} &
\scriptsize \ttfamily
My research spans mathematical physics, operator theory, and experimental particle physics, with a focus on spectral properties of the almost Mathieu operator, the structure of minimum phase preserving operators, and the inverse problems they present. I also contribute to the T2K long-baseline neutrino oscillation experiment, where I investigate CP violation, neutrino mixing parameters, and the sensitivity of extended experimental runs. My expertise lies in the interplay between mathematical theory and experimental analysis, particularly in quantifying and interpreting fundamental physical phenomena through rigorous mathematical and statistical frameworks. \\
\midrule
\multirow{1}{*}[-1.2em]{\rotatebox{90}{\textbf{Broad}}} &
\scriptsize \ttfamily
My research focuses on high-energy astrophysics, particularly X-ray spectroscopy of black holes and active galactic nuclei, with an emphasis on iron line diagnostics and accretion disk physics. I also study particle interactions in galactic environments, such as cosmic ray-induced emissions in molecular clouds and chemical abundances in AGN and starburst galaxies. Additionally, I explore fundamental physics topics including quantum Hall effects and pseudo-bosonic operator theory, reflecting a broader interest in both observational astrophysics and mathematical physics. \\
\bottomrule
\end{tabularx}
\end{table}

\subsection{Characterizing Profile Breadth}

NL profiles, which characterize a research's interests, can vary significantly in breadth. Some researchers have a wider scope of interests, while others focus on a more narrow topic. These differences can influence the performance of recommendation methods. 
To account for this, we automatically classify each profile's breadth and provide this classification as metadata. Specifically, we use a few-shot prompt (shown in Figure~\ref{fig:classification-prompt}) and take the majority vote from three distinct LLMs (Llama-70B, Gemini-2.5-Flash, GPT-4o) to ensure a robust classification that mitigates potential single-model biases~\citep{Balog:2025:SIGIR}. Table~\ref{tab:breadth_examples} shows examples of profiles from each category.

\begin{table}[t]
    \centering
    \caption{Summary statistics of the \dataset dataset.}
    \label{tab:dataset_summary}
    \begin{tabular}{lr} 
        \toprule
        \#Users & 1,000 \\
        \#Authored papers (min/median/max) & 10/20/260 \\
        \#Candidate items per user & 1,000 \\
        \#Ground truth papers per user (min/median/max) & 1/27/438 \\
        Profile length (words) & 117 $\pm$ 55 \\
        \phantom{xx} - Prompt A & 83 $\pm$ 10 \\
        \phantom{xx} - Prompt B & 150 $\pm$ 61 \\
        \#Narrow/Medium/Broad NL profiles & 679/256/65 \\
        \bottomrule
    \end{tabular}
\end{table}

\begin{figure*}[t!]
    \centering
    \begin{subfigure}[t]{0.48\textwidth}    
        \centering
        \vspace*{-\baselineskip}
        \includegraphics[width=0.75\linewidth]{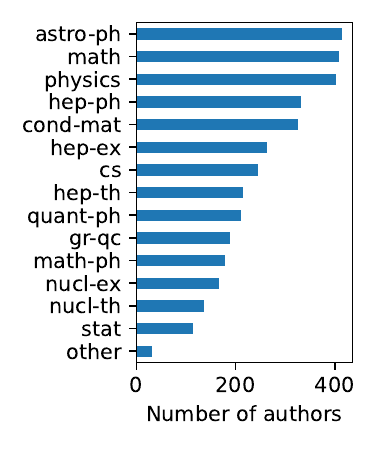}
        \vspace{-1.1\baselineskip}
        \caption{ArXiv categories (authors are counted in each category they have published in).}
        \label{fig:arxiv-categories}
    \end{subfigure}
    \hfill
    \begin{subfigure}[t]{0.48\textwidth}    
    \centering
    \vspace*{-\baselineskip}
        \includegraphics[width=0.75\linewidth]{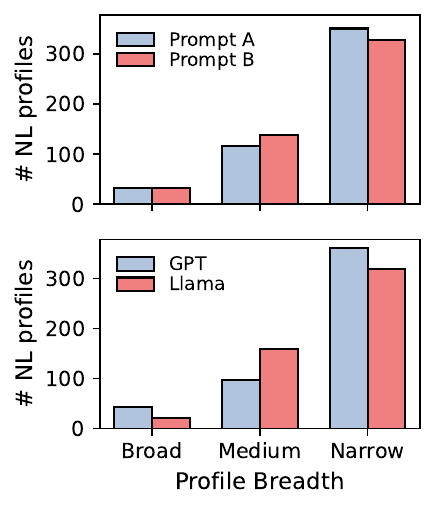}
        \vspace{-0.8\baselineskip}
        \caption{Profile breadth based on prompts and LLMs used.}
        \label{fig:breadth-distribution}
    \end{subfigure}
    \caption{Category and profile distributions in the \dataset dataset.}
\end{figure*}

\subsection{Dataset Statistics}
\label{sec:dataset:stats}

The key statistics for the \dataset dataset are summarized in Table~\ref{tab:dataset_summary}. 
With 1,000 users, the dataset supports robust statistical analysis, and the diversity in paper counts offers a realistic and varied set of user profiles.
The most frequently represented arXiv categories in which the users in our dataset publish are diverse fields of physics, mathematics, and computer science; see Figure~\ref{fig:arxiv-categories}.

We observe a noticeable difference in profile length across Prompts A and B, which will provide insights into the impact of prompting strategies (to be examined in Section~\ref{sec:exp:analysis}).

The majority of NL profiles are classified as narrow. Figure~\ref{fig:breadth-distribution} provides a further breakdown of profile breadth by the prompts and LLMs used. We find that the choice of prompt has a negligible effect on profile breadth. However, we observe a notable difference in the distributions based on the LLM used: GPT-4o tends to produce profiles that are either broad or narrow, while Llama yields a higher proportion of medium-breadth profiles.

\section{Methods for NL-based Recommendation}
\label{sec:baselines}

To establish strong performance baselines for NL-based recommendation, we benchmark a diverse range of existing methods. Our evaluation covers traditional sparse retrieval, modern dense retrieval, and state-of-the-art LLM-based reranking methods. The objective of this work is not to propose a novel method, but to systematically analyze the effectiveness of these established techniques within our specific problem setting, thereby providing a robust foundation for future research.

\vspace*{-0.5\baselineskip}
\subsubsection*{Sparse Retrieval Methods} We employ \textbf{BM25} as a traditional bag-of-words retrieval model, and \textbf{RM3} for query expansion using pseudo-relevance feedback.

\vspace*{-0.5\baselineskip}
\subsubsection*{Dense Retrieval Methods} 
We evaluate two main families of dense neural methods: dual-encoders and cross-encoders.

The first approach, using a \emph{dual-encoder architecture}, frames the recommendation task as a k-nearest neighbor (kNN) retrieval problem. The natural language user interest profile and the candidate items are independently embedded into a shared high-dimensional dense vector space. Subsequently, kNN search is employed to find the items closest to the user's profile vector. For this method, we use \textbf{SciBERT}~\citep{Beltagy:2019:EMNLP}, a BERT-based language model pre-trained on a large corpus of scientific text (768-dimensional embeddings) and we leverage Faiss~\citep{Douze:2024:arXiv} for efficient similarity search.

The second group utilizes \emph{cross-encoder models}. In contrast to dual-encoders, a cross-encoder processes the user profile and a candidate item jointly as a single input to output a relevance score for that pair. This allows for more precise relevance modeling but is computationally more intensive. We use these models to score each candidate document based on its relevance to the profile. Specifically, we selected several variants of BAAI's BGE (Beijing General Embedding) reranker models~\cite{Chen:2024:arXiv} due to their competitive performance across various retrieval tasks~\cite{Yang:2025:arXiv, Moreira:2024:arXiv}: \textbf{BGE-Large}\footnote{\url{https://huggingface.co/BAAI/bge-reranker-large}} and \textbf{BGE-v2-M3}\footnote{\url{https://huggingface.co/BAAI/bge-reranker-v2-m3}} (568 million parameters, based on XLM-RoBERTa-Large), and \textbf{BGE-v2-MiniCPM}\footnote{\url{https://huggingface.co/BAAI/bge-reranker-v2-minicpm-layerwise}} (2.72B parameters, based on MiniCPM). 

\vspace*{-0.5\baselineskip}
\subsubsection*{LLM-Based Reranking} We use Pairwise Relevance Prompting, \textbf{PRP}~\citep{Qin:2024:NAACL}, a state-of-the-art approach for predicting relative differences between a pair of items. This is shown to result in substantially better rankings than predicting absolute judgements in a pointwise manner. Due to the computational cost involved with LLM-based reranking, we employ it in a cascading manner by reranking the top-100 results from initial BM25 retrieval. Specifically, we utilize the most efficient \emph{sliding window} variant of PRP, which significantly reduces complexity ($O(n)$ for reranking $n$ items).    
    We employ three different LLM backbones: two public models, \textbf{Llama-3-8B}\footnote{\url{https://openrouter.ai/meta-llama/llama-3-8b-instruct}} and \textbf{Llama-3.3-70B},\footnote{\url{https://openrouter.ai/meta-llama/llama-3.3-70b-instruct}} as well as a proprietary model, \textbf{GPT-4o-mini}.\footnote{\url{https://openrouter.ai/openai/gpt-4o-mini}}

\vspace*{-0.5\baselineskip}
\subsubsection*{Ensemble} Additionally, we construct an ensemble retrieval method that combines the top-performing sparse, dense, and LLM-based models. Specifically, we apply the reciprocal rank fusion (RRF)~\cite{Cormack:2009:SIGIR} technique to combine results from RM3, BGE-v2-MiniCPM, and PRP-GPT-4o-mini.

\vspace*{-0.5\baselineskip}
\subsubsection*{Implementation} Sparse and kNN-SciBERT dense retrievals are implemented using Pyserini~\citep{Lin:2021:SIGIR}, while BGE dense models employ the FlagEmbedding library\footnote{\url{https://github.com/FlagOpen/FlagEmbedding}}. We use default parameter settings for BM25 and RM3, as well as for all BGE models. LLM-based reranking is implemented using the same tooling and API as NL profile generation; the prompt is included in the online repository.

\section{Results}
\label{sec:baselines:results}

This section reports on our experiments on natural language profile-based recommendation using the \dataset collection.

\begin{table}[t]
    \caption{Evaluation results. Highest scores within each family are italicized, highest overall scores are boldfaced.}
    \label{tab:results}
    \centering
    \begin{tabular}{lcccc}
        \toprule
        \textbf{Model} & \textbf{R@100} & \textbf{MAP} & \textbf{MRR} & \textbf{NDCG@10} \\      
        \midrule
        BM25 & 0.3491 & 0.1148 & 0.4661 & 0.2869 \\
        RM3 & \textit{0.3570} & \textit{0.1391} & \textit{0.5147} & \textit{0.3251} \\
        \midrule
        kNN-SciBERT & 0.1480 & 0.0232 & 0.2182 & 0.1019 \\
        BGE-Large & 0.2826 & 0.0783 & 0.3666 & 0.2072 \\
        BGE-v2-M3 & 0.3472 & 0.1152 & 0.4633 & 0.2763 \\
        BGE-v2-MiniCPM & \textbf{\textit{0.4203}} & \textit{0.1673} & \textit{0.5393} & \textit{0.3541} \\
        \midrule
        PRP-Llama-3 (8B) & 0.3491 & 0.1165 & 0.4774 & 0.2925 \\
        PRP-Llama-3.3 (70B) & 0.3491 & \textit{0.1423} & \textit{0.5378} & 0.3541 \\
        PRP-GPT-4o-mini & 0.3491 & 0.1405 & 0.5297 & \textit{0.3542} \\
        \midrule
        Ensemble & 0.4136 & \textbf{0.2163} & \textbf{0.6333} & \textbf{0.4481} \\
        \bottomrule
    \end{tabular}
\end{table}

Table~\ref{tab:results} summarizes the retrieval and reranking performance across sparse, dense, and LLM-based approaches in terms of traditional rank-based metrics: Recall@100, Mean Average Precision (MAP), Mean Reciprocal Rank (MRR), and NDCG@10. For significance testing we employ a paired t-test with Bonferroni correction at significance level of $\alpha = 0.05$.
Among the sparse methods, RM3  outperforms BM25 across all metrics (significant for all metrics except Recall@100), highlighting the enduring effectiveness of classical pseudo-relevance feedback~\citep{Lin:2019:SIGIRForum}. 
Surprisingly, most dense retrievers fail to outperform the sparse baselines. Within the dense retrieval family, BGE-v2-MiniCPM stands out, achieving the highest scores among all methods and surpassing both sparse and dense baselines (significantly so for all metrics except MRR against RM3). This demonstrates the potential of newer multitask-trained cross-encoder models to capture fine-grained semantic relevance more effectively than earlier dense architectures.
LLM-based rerankers, specifically the larger models PRP-Llama-3.3 (70B) and PRP-GPT-4o-mini, achieve performance comparable to the strongest dense retriever. These findings suggest that while dense and LLM-based methods are rapidly improving, sparse retrieval methods, particularly RM3, remain a formidable baseline for domain-specific search.

The ensemble model achieves the highest scores across all metrics except for Recall@100, significantly outperforming each of its constituent models (RM3, BGE-v2-miniCMP, and PRP-GPT-4o-mini). This aligns with previous findings in the scientific IR literature that even simple hybrid methods can outperform both sparse and dense approaches~\cite{Mandikal:2024:arXiv}. The ensemble's effectiveness also underscores the complementary nature of sparse, dense, and LLM-based signals.

\begin{figure}[t]
    \centering
    \vspace*{-\baselineskip}
    \includegraphics[width=\linewidth]{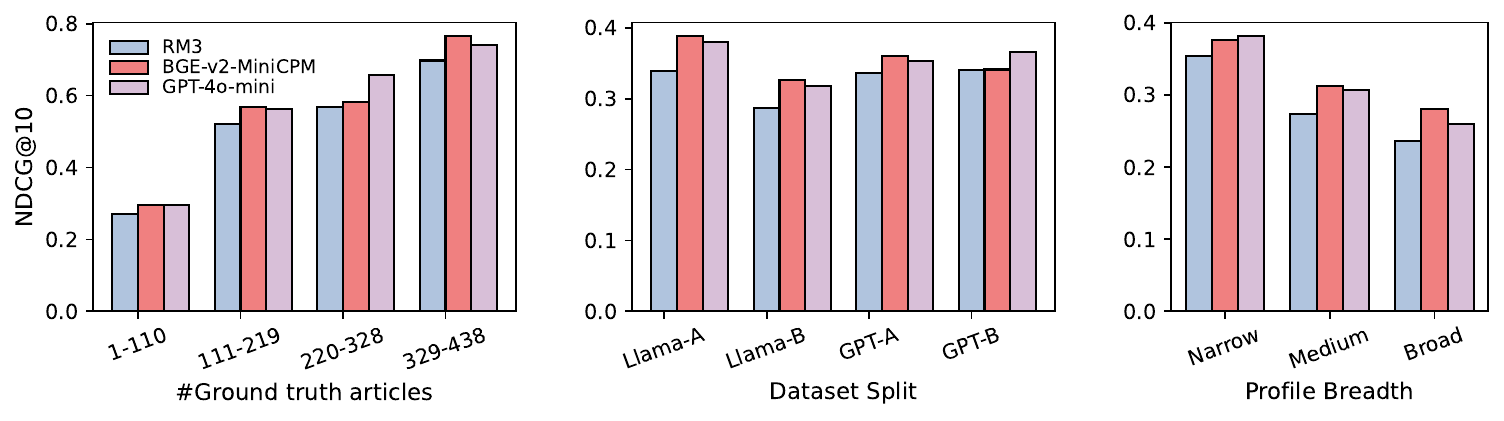}
    \vspace{-1.25\baselineskip}
    \caption{Performance comparison by number of ground truth articles (Left), by prompt-LLM combination used for NL profile generation (Middle) and by NL profile breadth (Right).}
    \label{fig:analysis}
\end{figure}

\subsection{Analysis}
\label{sec:exp:analysis}

To further analyze the impact of NL profile construction, we examine retrieval performance (in terms of NDCG@10) of our best-performing sparse (RM3), dense (BGE-v2-MiniCPM) and LLM-reranking (PRP-GPT-4o-mini) approaches.
Figure~\ref{fig:analysis} (Left) presents a breakdown based on the number of ground truth items. Unsurprisingly, we find that performance improves with more ground truth items available. 
Figure~\ref{fig:analysis} (Middle) breaks down performance based on the specific combination of prompt and LLM used for profile generation. 
We find that Prompt A (over Prompt B) and GPT (over Llama) yield more effective profiles, however, none of these differences is statistically significant.
Crucially, Figure~\ref{fig:analysis} (Right) reveals that broader profiles tend to degrade performance. This is likely because their diverse topics are difficult to represent effectively in a single query, resulting in a lack of topical focus that challenges sparse retrievers, dense retrievers, and LLM-rerankers alike. This finding underscores the need for more sophisticated methods capable of interpreting diverse user interests.

\begin{figure}[t]
    \centering
    \vspace*{-\baselineskip}
    \includegraphics[width=\linewidth]{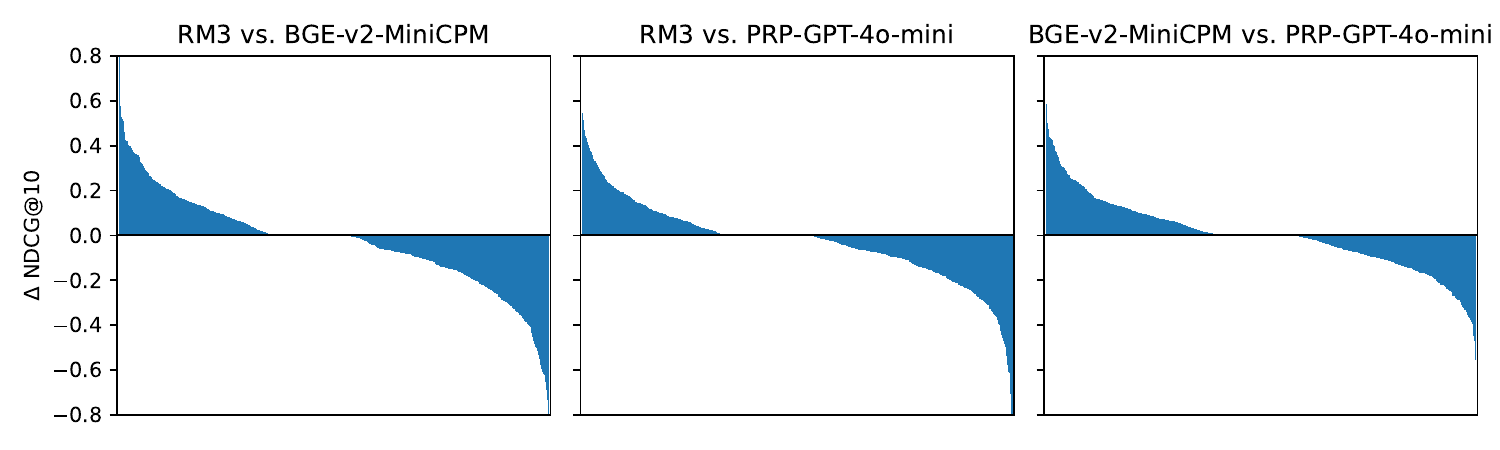}
    \caption{Per-author NDCG@10 differences between retrieval models. Positive values indicate higher scores for the first model in each comparison.}
    \label{fig:diffs}
\end{figure}

To investigate model complementarity,  Figure~\ref{fig:diffs} plots per-author NDCG@10 differences between the best-performing models from each family. 
While one model may outperform another on aggregate, the plot reveals substantial variability at the author level, where the globally weaker model often achieves superior results. 
This demonstrates that different retrieval approaches capture complementary relevance signals and that no single model is universally optimal. Consequently, this variability highlights a clear opportunity for hybrid models that combine diverse retrieval paradigms to achieve more robust and personalized recommendation performance.

\section{Conclusion}
\label{sec:concl}

We introduced \dataset, a novel dataset for advancing NL-based scholarly recommendation. We detailed its construction process, and established baselines for the core task of \emph{NL profile-based recommendation} using a range of sparse, dense, and LLM-based approaches. Our experiments reveal that different retrieval paradigms capture complementary relevance signals, and a simple ensemble of top models confirms that substantial headroom for improvement remains. These findings establish \dataset as a valuable testbed for developing novel methods, particularly hybrid architectures designed to fuse these complementary signals and models that can robustly interpret diverse user profiles.

Our work also opens several exciting avenues for the task of \emph{NL profile generation}.
The construction process involves various parameters and prompting strategies that can be further explored, such as the ratio of papers used for profile creation versus evaluation and the specific instructions given to the LLM. Furthermore, validating the generated NL profiles through human evaluation would provide crucial insights into their quality and fidelity to actual user interests, guiding the development of more effective profile generation techniques.

\small{\subsubsection{\ackname} This research was supported by the Norwegian Research Center for AI Innovation, NorwAI (Research Council of Norway, project number 309834).}

\small{\subsubsection{Disclosure of Interests.}
The authors have no competing interests to declare that are relevant to the content of this article.}

\bibliographystyle{splncs04nat}
\bibliography{arxiv2025-nlprofiles}

\end{document}